\documentclass[mathleft,fleqn]{an}

\usepackage{amssymb}
\usepackage{graphicx}
\usepackage[varg]{txfonts}
\usepackage{textcomp}

\usepackage{hyperref}
\hypersetup{%
    final=true,
    pageanchor=true,
    colorlinks=true,
    breaklinks=true,
    linkcolor=blue,
    citecolor=blue,
    urlcolor=blue,
    pdfpagemode=UseNone,
    pdftitle={Solar Physics at the Einstein Tower},
    pdfauthor={C. Denker et al.},
    pdfsubject={solar instrumentation},
    pdfkeywords={history and philosophy of astronomy,
        Sun: photosphere,
        Sun: magnetic fields,
        Sun: flares,
        techniques: spectroscopic,
        telescopes}}

\usepackage[modulo, switch]{lineno}

\usepackage{natbib}
\bibpunct{(}{)}{;}{a}{}{,}

\newcommand\mytextdied{\raisebox{0.5ex}{\footnotesize\textdied}}

\definecolor{com}{rgb}{0.698039, 0.0941176, 0.133333}

\begin{document}


\Pagespan{1}{}
\Yearpublication{2016}
\Yearsubmission{2016}
\Month{0}
\Volume{999}
\Issue{0}
\DOI{asna.201400000}

\title{Solar Physics at the Einstein Tower}

\author{%
    C.\ Denker\inst{1}\fnmsep\thanks{Corresponding author: 
    \href{mailto:cdenker@aip.de}{cdenker@aip.de}},
    C.\ Heibel\inst{1,2},
    J. Rendtel\inst{1},
    K.\ Arlt\inst{1},
    H.\ Balthasar\inst{1},
    A.\ Diercke\inst{1,3},\\
    S.~J.\ Gonz\'alez Manrique\inst{1,3},
    A.\ Hofmann\inst{1},
    C.\ Kuckein\inst{1},
    H. \"Onel\inst{1},
    V.~Senthamizh Pavai\inst{1,3},\\
    J.\ Staude\inst{1}, \and
    M.\ Verma\inst{1}}

\institute{%
    Leibniz-Institut f{\"u}r Astrophysik Potsdam (AIP),
    An der Sternwarte 16, 
    14482 Potsdam, Germany
\and
    Humboldt-Universit\"at zu Berlin,
    Unter den Linden 6,
    10099 Berlin, Germany
\and 
    Universit{\"a}t Potsdam, 
    Institut f{\"u}r Physik and Astronomie, 
    Karl-Liebknecht-Stra{\ss}e 24/25,
    14476 Potsdam, Germany}

\titlerunning{History of Solar Physics at Potsdam}
\authorrunning{C.\ Denker et. al.}

\received{\today}
\accepted{XXXX}
\publonline{XXXX}


\abstract{The solar observatory Einstein Tower (\textit{Einsteinturm}) at  the 
Telegrafenberg in Potsdam is both a landmark of modern architecture and an 
important place for solar physics. Originally built for high-resolution 
spectroscopy and measuring the gravitational redshift, research shifted over the 
years to understanding the active Sun and its magnetic field. Nowadays, 
telescope and spectrographs are used for research and development, i.e., testing 
instruments and in particular polarization optics for advanced instrumentation 
deployed at major European and international astronomical and solar telescopes. 
In addition, the Einstein Tower is used for educating and training of the next 
generation astrophysicists as well as for education and public outreach 
activities directed at the general public. This article comments on the 
observatory's unique architecture and the challenges of maintaining and 
conserving the building. It describes in detail the characteristics of 
telescope, spectrographs, and imagers; it portrays some of the research and 
development activities.}
\maketitle

%
%

\section{Introduction}

\begin{figure}[t]
\includegraphics[width=\columnwidth]{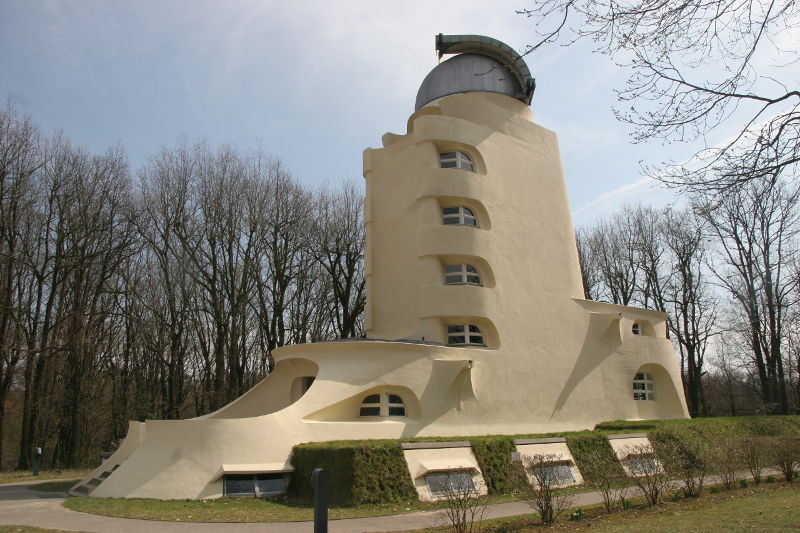}
\caption{North-west view of the Einstein Tower in its original light 
    yellow-ochre color scheme. The dome is opened for solar observations. The
    grass covered basement contains optical laboratory, spectrograph chamber, 
    and workshops.}
\label{FIG01}
\end{figure}

The $12^\mathrm{th}$ Potsdam Thinkshop in October 2015 was dedicated to the 
theme ``The Dynamic Sun -- Exploring the Many Facets of Solar Eruptive Events''. 
The Einstein Tower served as the background image of the conference poster 
(Fig.~\ref{FIG01}), and the meeting concluded with a visit to the Einstein Tower 
highlighting the major achievements of this observatory, its remarkable 
architecture, and its contemporary use as a laboratory for instrument 
development and as a place for training the next generation of astrophysicists. 
The great interest expressed by the participants motivated us to write this 
article from the perspective of scientists who were and still are active at the 
solar observatory Einstein Tower. 

The Astrophysical Observatory Potsdam (AOP) was founded in 1874 as the first 
observatory carrying the term astrophysics in its name \citep{Staude1991}. This 
marks the transition from astronomy focused on external characteristics of 
celestial objects to astrophysics as a theory-driven science exploring the 
physical nature of these bodies. The Telegrafenberg in Potsdam was named after 
the Prussian semaphore system (built in 1832) for optical communication between 
Berlin and Koblenz. Besides AOP, the historical science park comprises several 
institutes such as the Magnetic Observatory (1889), the Meteorological 
Observatory (1890), and the Geodetic Institute (1892). A location to the South 
of the Great Refractor (1899) was chosen as the construction site of the 
Einstein Tower, where erection of the tower started in the summer of 1920, 
finishing the shell construction already one year later. The construction faced 
economically challenging times (postwar reconstruction and hyperinflation). 
Interior construction and outfitting of the laboratory and work spaces commenced 
in 1922/23. The telescope optics and its support structure were manufactured by 
Carl Zeiss in Jena starting in 1919. The telescope was installed in 1924, and 
first observations were carried out on 1924 December~6 \citep{Jaeger1986}. The 
Einstein Tower thus complemented the broad spectrum of research concentrated on 
the Telegrafenberg, and in particular continued and expanded solar physics 
research, which was always an integral part of AOP's astrophysical research 
program \citep{Staude1995}. A detailed description of the architectural 
achievements and of the historical background is presented in the monographs by 
\citet{Eggers1995} and \citet{Wilderotter2005}.

The Einstein Tower is a unique landmark of expressionist, utopian, and symbolic 
architecture at the beginning of the 20$^\mathrm{th}$ century -- initiated by 
the astronomer Erwin Finlay Freundlich ($^\ast$1885\,--\,\mytextdied 1964), 
designed by the architect Erich Mendelsohn ($^\ast$1887\,--\,\mytextdied 1953), 
and supported by Albert Einstein ($^\ast$1879\,--\,\mytextdied 1955). Their 
interaction was driven by the ambition to find observational proof of the 
general theory of relativity. This endeavor also received support by Karl 
Schwarzschild ($^\ast$1873\,--\,\mytextdied 1916), director of AOP, who himself 
contributed to the general theory of relativity with the first exact solutions 
to the field equations. This intriguing paradigm shift from astronomy to 
astrophysics, marking a new era of physics -- the confluence of quantum theory 
and relativity, was explored by \citet{Hentschel1992} in his monograph from the 
perspective of the history of science.

\begin{figure}[t]
\includegraphics[width=\columnwidth]{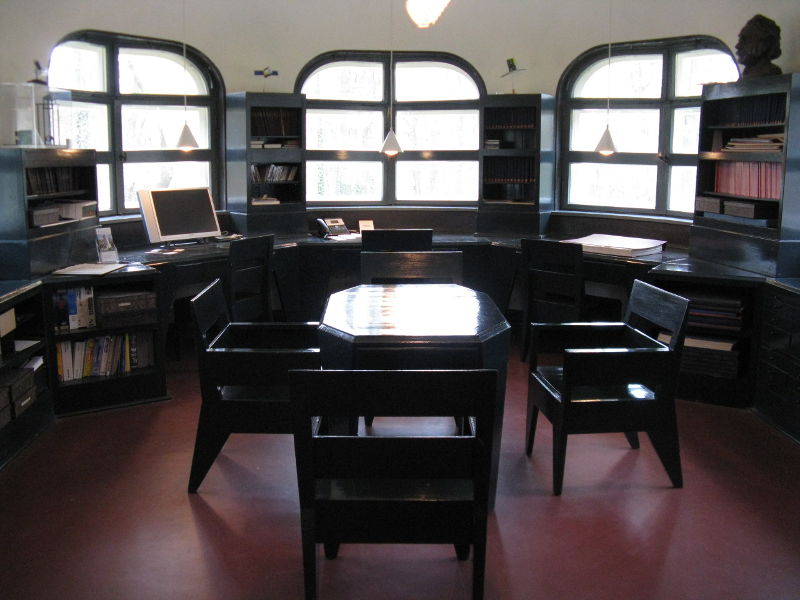}
\caption{A small library and the photographic plate archive of the Einstein 
    Tower are contained in the meeting room on the first floor. Composition of 
    interior walls and furnishing were conceptualized in its entirety by Erich 
    Mendelsohn, who also designed the furniture.}
\label{FIG02}
\end{figure}

Einstein released a preliminary version of his general theory of relativity in 
1911, which expands the special theory of relativity to include accelerated 
frames of reference and the equivalence principle of gravitational and inertial 
mass. He pointed out three effects that are in principle accessible to 
astronomical observations \citep{Jaeger1986}: (1) rotation of apsidal nodes of 
planetary orbits (known since about 1850 for Mercury), (2) bending of light when 
passing heavy objects such as stars and galaxies, and (3) the gravitational 
redshift. The confirmation of the second effect, during an eclipse expedition by 
Sir Arthur Eddington and Frank Watson Dyson in 1919, was the starting signal and 
impetus to raise funding for the Einstein Tower. The new solar observatory was 
financed in equal parts by the state of Prussia and the Albert Einstein 
Foundation of the German Industry. The foundation's board of trustees included 
among others Einstein as the chairman, Freundlich as the head of the institute, 
and  the director of AOP. The board convened regularly in the meeting room of 
the Einstein Tower, which was outfitted with furniture also designed by 
Mendelsohn (Fig.~\ref{FIG02}). Board meetings were concerned with the operation 
of the Einstein Tower, i.e., finances, acquisition of new instruments, and 
recurrent structural damages to the building.

Freundlich as head of the Einstein Institute was in charge of the Einstein Tower 
from 1920 to 1933. With the take-over of power by the National Socialist Party 
in 1933, Freundlich emigrated to Turkey and  Einstein's name was removed from 
the institute by renaming it into Institute for Solar Physics, and soon 
thereafter the observatory was absorbed and integrated in AOP. The monography by 
\citet{Sailer2007} recounts the history of solar research in Germany from 
1939\,--\,1949 and its entanglement with the German Luftwaffe. After World 
War~II, the Einstein Tower belonged to the German Academy of Sciences (in 1946), 
the Central Institute for Solar-Terrestrial Physics/Heinrich Hertz Institute (in 
1967), and the Central Institute for Astrophysics (in 1983). Following the 
German reunification, the solar observatory became part of the Astrophysical 
Institute Potsdam (1992), which was renamed to Leibniz Institute for 
Astrophysics Potsdam (AIP) in 2011.

%
%

\section{Preserving the Einstein Tower}

The mixed construction of the Einstein Tower consisting of bricks, concrete, and 
rendering is the underlying cause for a recurring damage pattern. The concrete 
quality and iron reinforcements do not conform to current standards, and no 
efficient way to remove rain water and humidity was available leading to 
fissures in the rendering coat, which facilitate an easy entry of rain water and 
moisture. This caused damages after freezing temperatures and corrosion of the 
iron reinforcements (structural members) above windows and throughout the 
building structure \citep{Pitz1996}. 

The first major repairs in 1927 because of widespread moisture penetration 
commenced under the supervision of Mendelsohn himself leading to a massive 
addition of sheet metal (window sills and roofing) significantly changing the 
character of the building. These measures, however, were not very effective 
necessitating major repairs about every 5\,--\,10 years. Regrettably, the 
history of the Einstein Tower is also a history of recurring damages and endless 
renovations. The second major overhaul in 1940/41 was driven by fungal 
infestation that damaged the prism spectrograph in 1937. Towards the end of 
World War~II, a blockbuster detonation during an air-raid in proximity to the 
Einstein Tower severely damaged entrance, dome, windows, doors, and even 
interior walls. However, immediate repairs in September 1946 allowed solar 
observations to continue soon thereafter. Major repairs of the building followed 
in 1950, 1958, 1964, 1974\,--\,1978, and 1984. By the beginning of the 1990ies 
the condition of the Einstein Tower deteriorated to the point, where the 
continued existence of the building was endangered \citep{Pitz1996}.

\begin{figure}[t]
\includegraphics[width=\columnwidth]{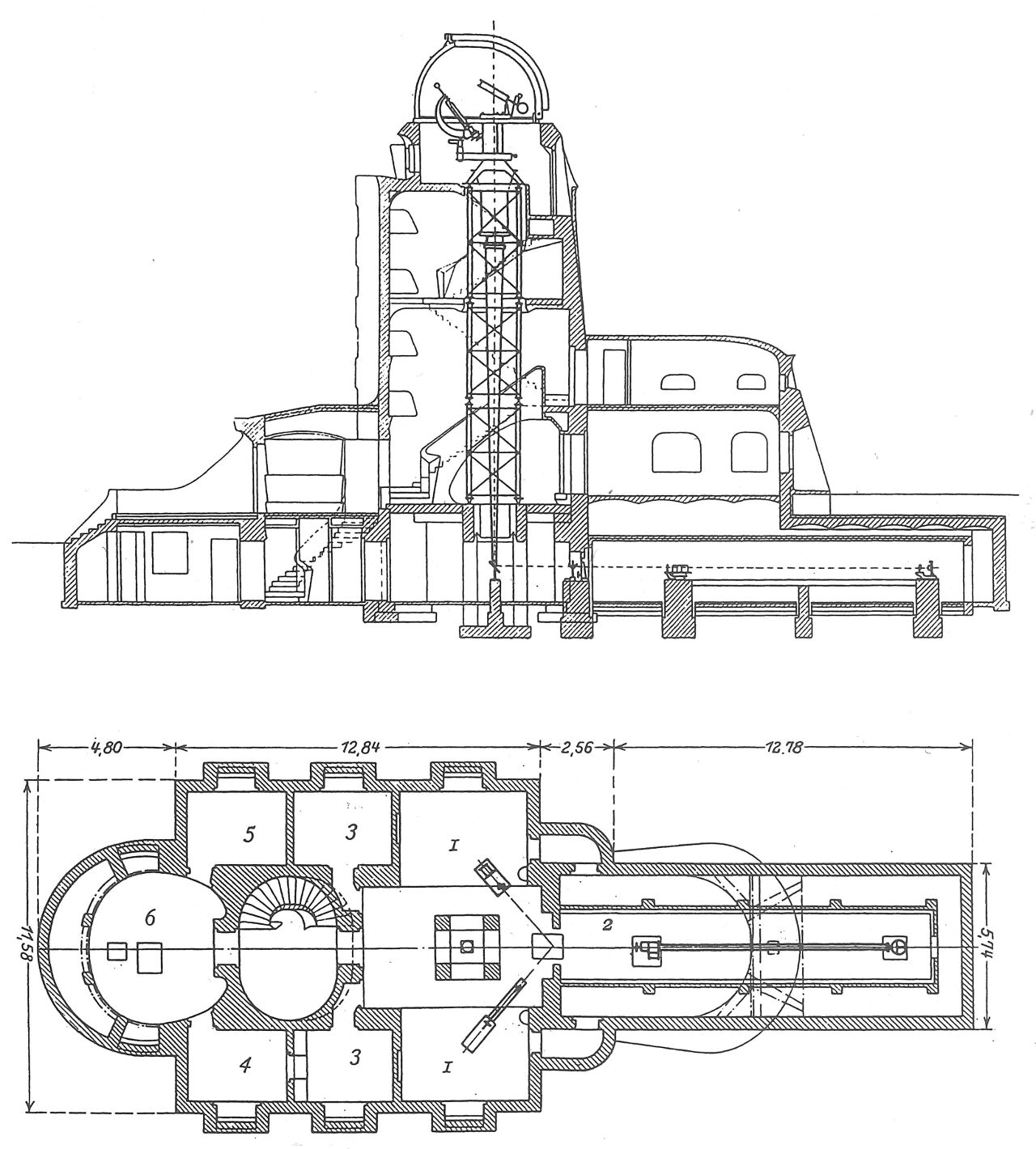}
\caption{Vertical section of the Einstein Tower (\textit{top}) and horizontal 
    section of its basement (\textit{bottom}) with the historical floor plan: 
    (1) laboratory with turning mirror, (2) thermally stabilized spectrograph 
    chamber, (3) workshops, (4) dark room, (5) high-voltage battery, and (6) 
    micro-photometer room \citep[see Fig.~2 in][]{Freundlich1927}.}
\label{FIG03}
\end{figure}

Intensive research into the origin and causes of the recurring damage pattern 
preceded the latest major repair, which was made possible by generous support of 
the W\"ustenrot Foundation (covering two-thirds of about 1.5 Million Euros) and 
was completed in 1999. The restoration was carried out in close collaboration 
with State Office for Monument Protection (Landesamt f\"ur Denkmalpflege) and 
included: renewal of the rendering coat conserving the original rendering 
whenever possible, return to an ochre-colored coat of paint for the entire 
facade, removal of the ineffective sheet metal installed in 1927, returning the 
dominant entrance terrace to its original appearance, and elimination of 
dampness and moisture damage in the basement. The guiding principles of the 
renovation and restoration were to preserve the artistic and corporeal substance 
of the Einstein Tower as much as possible, to facilitate easy access and 
procedures for future repairs, and to safeguard the building while upholding its 
function as a place for astrophysical research. At the end of the renovation it 
became evident that continuous monitoring and repairs will be required in the 
future \citep{Pitz2006a}. This led to a long-term conservation plan 
\citep{Pitz2006b} because causes for damage could not be completely removed but 
only mitigation plans were put into place.

\begin{figure}[t]
\includegraphics[width=\columnwidth]{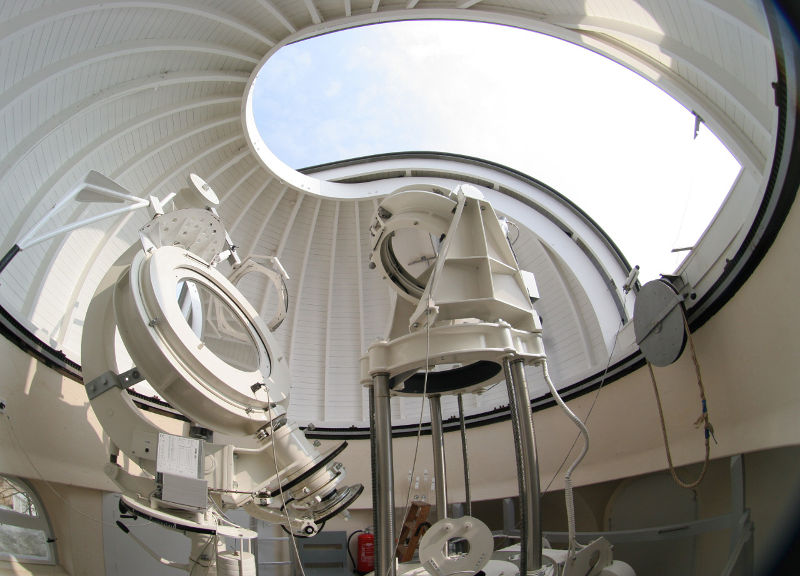}
\caption{Coelostat system inside the metal-plated wooden dome. The dome rotation
    is motorized but the shutter has to be opened manually. The boom over the
    first coelostat mirror carries a large optical mount for polarization 
    calibration. A mechanism to change the telescope aperture is located below 
    the second coelostat mirror, which sends the light vertically into the 
    tower.}
\label{FIG04}
\end{figure}

A fire during the restoration of the Einstein Tower on 1998 January~6 led to 
significant damages to the wooden support structure and the telescope itself. 
The mechanical mounts of the coelostat mirrors and the entrance aperture were 
cleaned of soot and other remnants of the fire. Cleaning of the 60-cm doublet 
lens required a careful evaluation of potential surface damages, which 
fortunately had not occurred, so that after a special cleaning procedure, the 
lens could be restored to its previous performance. On 1999 July~27, the 
Einstein Tower resumed operations \citep{Balthasar2000b}.

In the 2015 competition of the Getty Foundation's ``Keeping It Modern'' 
initiative the Einstein Tower received a planing and conservation grant. Several 
major tasks are currently carried out supported by this grant: (1) validate that 
the cyclic maintenance plan is effective and sustainable, (2) identify an 
efficient solution to prevent dampness and moisture damage in the subterranean 
laboratory, which houses the high-resolution spectrograph with its sensitive 
optics and gratings, (3) install a better heating and temperature control system 
for the basement using modern, energy efficient systems, and (4) investigate the 
thermally induced stresses in the hull of the Einstein Tower using simulations 
to clearly identify weaknesses in the building structure, to reveal hidden 
causes for damages, and to guide future mitigation strategies.

\begin{figure}[t]
\includegraphics[width=\columnwidth]{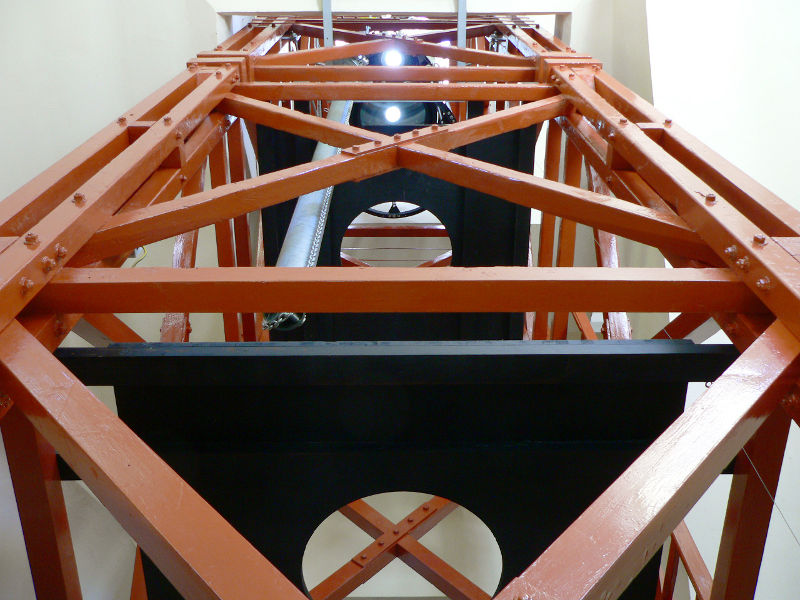}
\caption{The interior tower consists of a wooden scaffolding supporting the
    coelostat assembly. The wooden structure dampens vibrations and is 
    additionally erected on its own foundation. Black apertures prevent stray
    light from entering the optical path.}
\label{FIG05}
\end{figure}

%
%

\section{Telescope}

The tower telescopes at the Mt.\ Wilson Observatory in California served as an 
inspiration for the Einstein Tower \citep{Jaeger1986}, which was the first solar 
tower telescope built in Europe (vertical and horizontal sections of the 
Einstein Tower are presented in Fig.~\ref{FIG03}). The telescope is protected by 
a wooden dome with a diameter of 4.5~m. The wood paneling prevents condensation 
and the build-up of dew. The exterior of the dome is protected by a sheet metal 
cladding against the forces of the elements. Two plane coelostat mirrors with an 
aperture of 60~cm catch the sunlight at a height of 15~m above ground and direct 
the light beam vertically into the tower (Fig.~\ref{FIG04}). The main mirror has 
a parallactic mount, which can be rotated in azimuth preserving its parallactic 
orientation. This prevents vignetting by the shadow casted by the secondary 
mirror and its mount. Height and tilt of the secondary mirror are adjustable to 
accommodate the Sun's changing zenith distance over the course of a year and the 
azimuth angle of the main mirror (north corresponds to the normal position). The 
original silver-coated 85-cm coelostat mirrors were replaced in the mid-1950ies 
by 60-cm aluminum-coated glass mirrors. Since 1993 Zerodur mirrors are installed 
with much improved heat expansion properties minimizing deformations under heat 
load. Only a single optical element, i.e., a doublet objective lens with a 
diameter of $D = 60$~cm and a focal length of $f \approx 1400$~cm, forms a solar 
image with $d_{\scriptscriptstyle\bigodot} \approx 13$~cm diameter in the 
optical laboratory, which is located in the basement of the building. Here, 
another plane folding mirror with a diameter of 45~cm about 3~m before the 
primary focus directs the light beam into the horizontal. The 
diffraction-limited resolution in the primary focal plane is $\alpha = \lambda / 
D = 0.17\arcsec$ at 500~nm, and the plate scale is $s = 14.73\arcsec$~mm$^{-1}$.

\begin{figure}[t]
\centering
\includegraphics[height=\columnwidth]{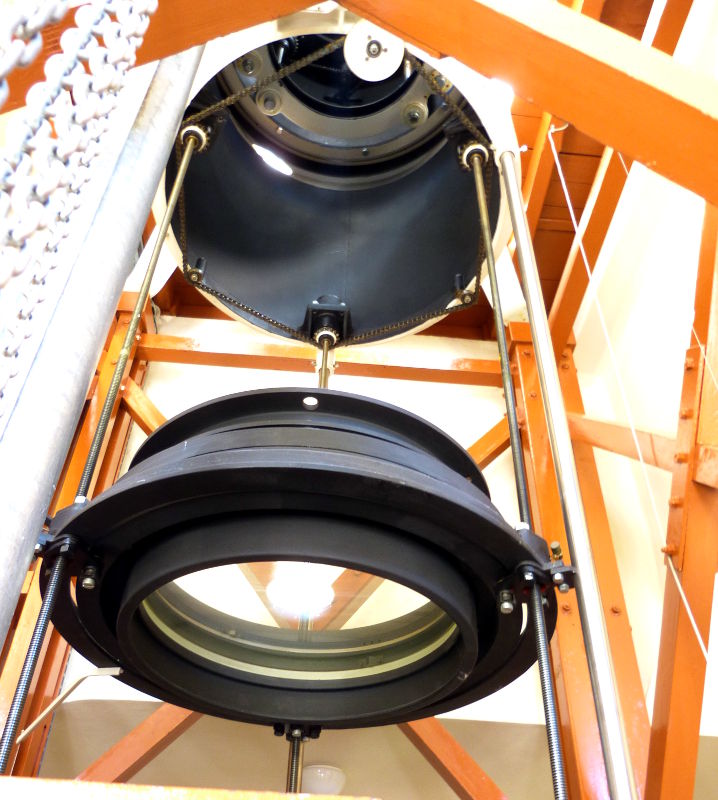}
\caption{Mounting of the 60-centimeter doublet lens with gearing mechanism to 
    adjust the focus position.}
\label{FIG06}
\end{figure}

An aperture stop is situated just below the second coelostat mirror (white 
donut-shaped object at the bottom of Fig.~\ref{FIG04}), and it is operated via 
a pulley from the basement laboratory to select an aperture with a diameter of 
10, 15, or 30~cm. Removing the lid with the apertures and pulley from the 
telescope structure yields the full aperture with a diameter of 60~cm. The 
smaller apertures are typically used for taking images facilitating higher 
contrasts when matching seeing- and diffraction-limited resolution. The larger 
apertures are more suitable for spectroscopic observations, even though highly 
efficient CCD detectors also deliver good results with the smaller apertures.

The telescope is installed on two stacked wooden towers (Fig.~\ref{FIG05}) with 
a separate foundation to insulate the telescope from the wind buffeting the 
building and vibrations caused by people moving inside the building or outside 
traffic. Ball and socket joints between foundation, wooden towers, and telescope 
mount further reduce propagation of vibrations. The low heat expansion 
coefficient of wood also minimizes thermal expansion and thus focus drifts over 
the day. The lens mount is attached to three threaded rods for focusing 
(Fig.~\ref{FIG06}). The vertical lifting distance is more than a meter in both 
directions, conveniently placing the focal plane where required for different 
set-ups in the optical laboratory. This is needed in particular, when inserting 
the magnification system for direct viewing of the 50-cm diameter solar image.


\subsection{Spectrograph}

\begin{figure}[t]
\includegraphics[width=\columnwidth]{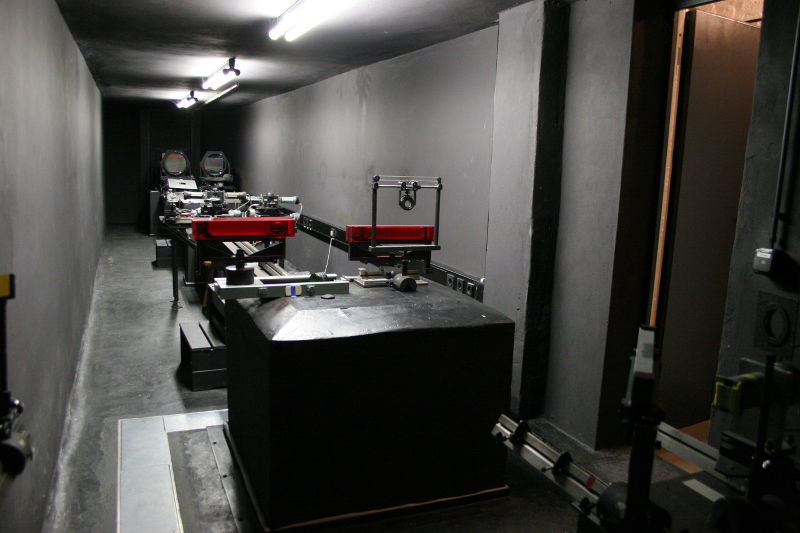}
\caption{Interior of the spectrograph chamber with the twin high-resolution
    spectrographs. Large-format gratings are located in the back behind 
    collimating lenses used in double path. An optical bench on a separate
    foundation supports the spectrograph optics and facilitates insertion of
    additional optical elements. Turning mirrors (one in the foreground and two
    with red protective plastic covers) redirect the light to (polarization)
    optics or CCD cameras.}
\label{FIG07}
\end{figure}

The two spectrographs of the Einstein Tower are located in the southern part of 
the basement, where the upper part (about 1~m, see Fig.~\ref{FIG07}) of the room 
is still above ground but below a grass-covered surface (see Figs.~\ref{FIG01} 
and \ref{FIG03}). The purpose of the sod is to minimize diurnal temperature 
changes inside the spectrograph chamber because of its exposed sunlit placement. 
This technique is nowadays often encountered in green flat-roofed buildings. 
Another measure to provide a stable thermal environment, which is required for 
high resolution spectroscopy, is the separate spectrograph chamber inside the 
basement. The chamber is surrounded on three sides by a corridor, which gives 
access to camera ports and offers some storage space. The air between interior 
and exterior walls and above the roof of the spectrograph chamber provides the 
thermal insulation and acts as a barrier for moisture penetrating the exterior 
walls. Electric dehumidifiers remove atmospheric moisture and keep the humidity 
at a constant level in the gallery. The concept of a nested spectrograph chamber 
was also included in the much more elaborate design of PEPSI 
\citep{Strassmeier2015}, however, with a much higher stability for temperature 
and humidity. The interior of the spectrograph chamber is painted in a matt 
black finish to minimize stray light (see Fig.~\ref{FIG07}). The spectrographs 
rest on their own foundation, which is separated from the rest of the building 
to reduce vibrations and mechanical shocks. A long optical rail, again on its 
own foundation, is placed in front of the main spectrograph so that polarization 
optics and measuring devices can be mounted.

The spectrograph chamber houses two high-resolution spectrographs with different 
gratings. The gratings have dimensions of 420\,mm $\times$ 320\,mm and 320\,mm 
$\times$ 220\,mm, respectively. The corresponding grating constants are $1 / 
d_{1,\,2} = 632$ and 594 lines mm$^{-1}$ leading to a total number of lines 
$N_{1,\,2} \approx 265\,000$ and 190\,000. The  spectral resolution is thus 
$\Re_{1,\,2} = m N_{1,\,2} \approx \lambda / \Delta\lambda_{1,\,2} = 
1\,000\,000$ and 760\,000 for observations in the fourth order ($m=4$), which 
yields $\Delta \lambda_{1,\,2} \approx$ 0.59 and 0.83~pm at an observed 
wavelength of $\lambda = 630.2$~nm. The grating equation $m \lambda = d 
(\sin\alpha + \sin\beta)$, where $\alpha$ and $\beta$ are the angles of 
incidence and reflection, simplifies in the Littrow configuration of the 
spectrographs ($\alpha \simeq \beta$) to $m \lambda = 2d \sin\beta$. Thus, for 
blaze angles $\beta_{1,\,2} = 52.8^\circ$ and $48.5^\circ$ the spectral region 
at $\lambda = 630.2$~nm is observed in the fourth order. Each spectrograph uses 
a 35-cm diameter lens made by Zeiss in double path, i.e., they serve at the same 
time as collimator and camera lenses. The focal length of the lenses is 
identical ($f_\mathrm{S} = 1200$~cm) and roughly matches the focal length of 
the telescope $f_\mathrm{T} = 1400$~cm. The linear dispersion for both 
spectrographs is given as
\begin{equation}
\left(\frac{\Delta\lambda}{\mathrm{d}x}\right)_{1,\,2} = 
    \frac{d_{1,\,2}\cos\beta_{1,\,2}}{m f_\mathrm{S}} = 
    19.9~\mathrm{and}~23.3~\mathrm{pm}~\mathrm{mm}^{-1}.
\label{EQN01}
\end{equation}

Since early 2016, an Alta F9000 full frame CCD camera manufactured by Apogee 
(now part of Andor Technology) was installed at the Einstein Tower replacing 
outdated CCD cameras. The main area of application is spectropolarimetry using 
the high-resolution spectrographs. Thus, a large-detector format and a good 
sensitivity were the selection criteria leading to the acquisition of this 
camera with a KAF-090000 front-illuminated sensor. The manufacturer 
specifications \citep{Andor2016} provide the following characteristics: The 
detector has $3056 \times 3056$ pixels, and the pixel size is 12~$\mu$m $\times$ 
12~$\mu$m. Therefore, the total imaging area is  36.7~mm $\times$ 36.7~mm. The 
full well capacity is 94\,000~e$^-$ with a dark current of 0.07~e$^-$ 
pixel$^{-1}$ s$^{-1}$ and a read noise of 16.1~e$^-$ at 2.9 MHz. The readout is 
relatively slow with 0.3 frames s$^{-1}$ but still sufficient for 
spectropolarimetric observations. Thermoelectric in combination with forced air 
cooling stabilizes the detector temperature to 45~$^\circ$C below ambient. The 
dynamic range is 75.3 dB (5840:1) so that the 16-bit digitization is more than 
required. The spectral coverage of the detector is 360\,--\,1080~nm with a 
maximum quantum efficiency of 70\% at 580~nm. The quantum efficiency is 34\% at 
400~nm, which facilitates observations of the important chromospheric absorption 
lines Ca\,\textsc{ii}\,H\,\&\,K. In the near-infrared, the quantum efficiency 
drops to below 10\% at 980~nm. Equation~\ref{EQN01} yields a spectral coverage 
of 0.73~nm in the order $m=4$ for the 630.2~nm spectral region and a dispersion 
of 0.24~pm pixel$^{-1}$.

\begin{figure}[t]
\includegraphics[width=\columnwidth]{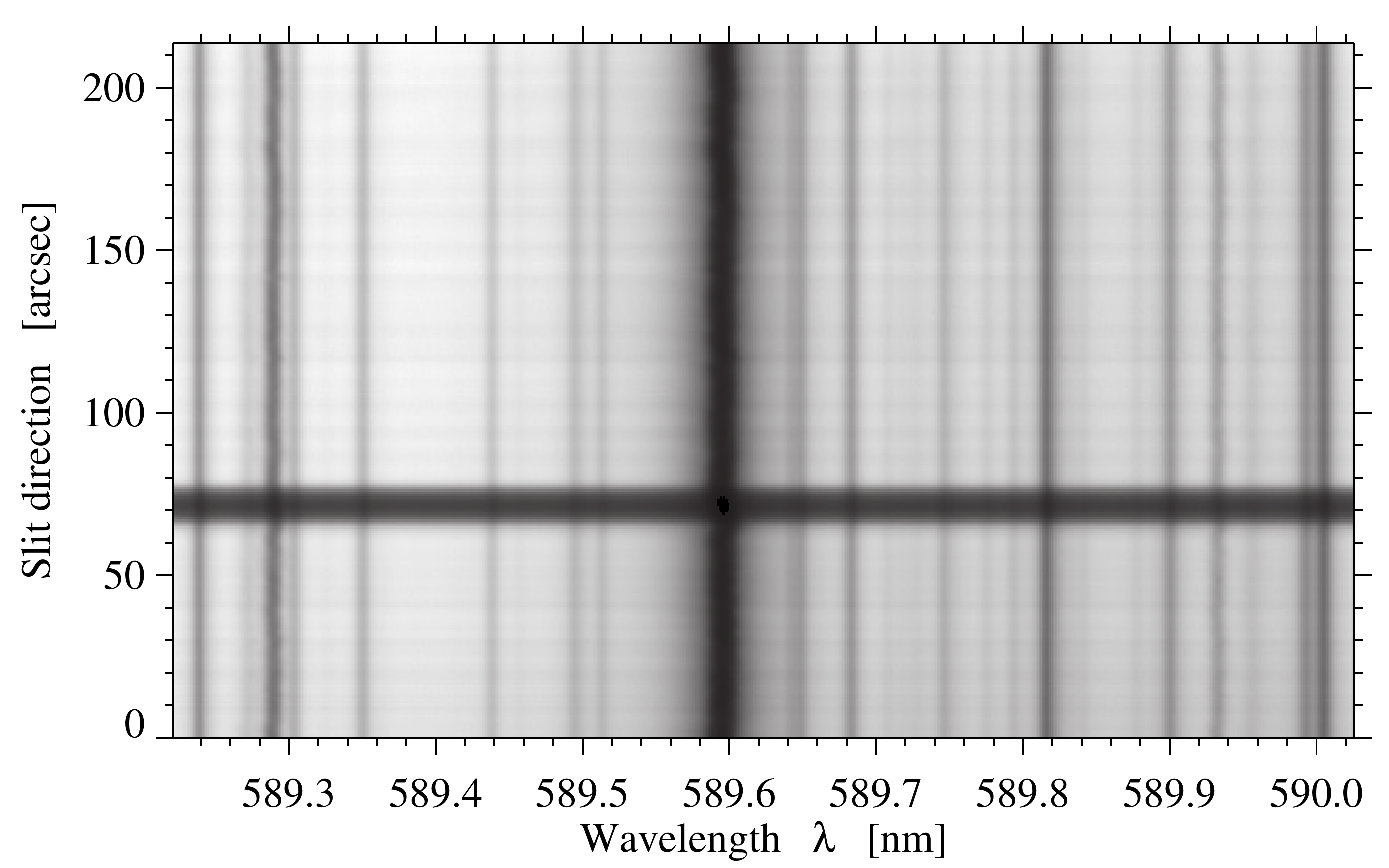}
\caption{Spectrum of the chromospheric Na\,D$_1$ 589.6~nm line observed during 
    the Mercury transit on 2016 May~9.}
\label{FIG08}
\end{figure}

\begin{figure}[t]
\includegraphics[width=\columnwidth]{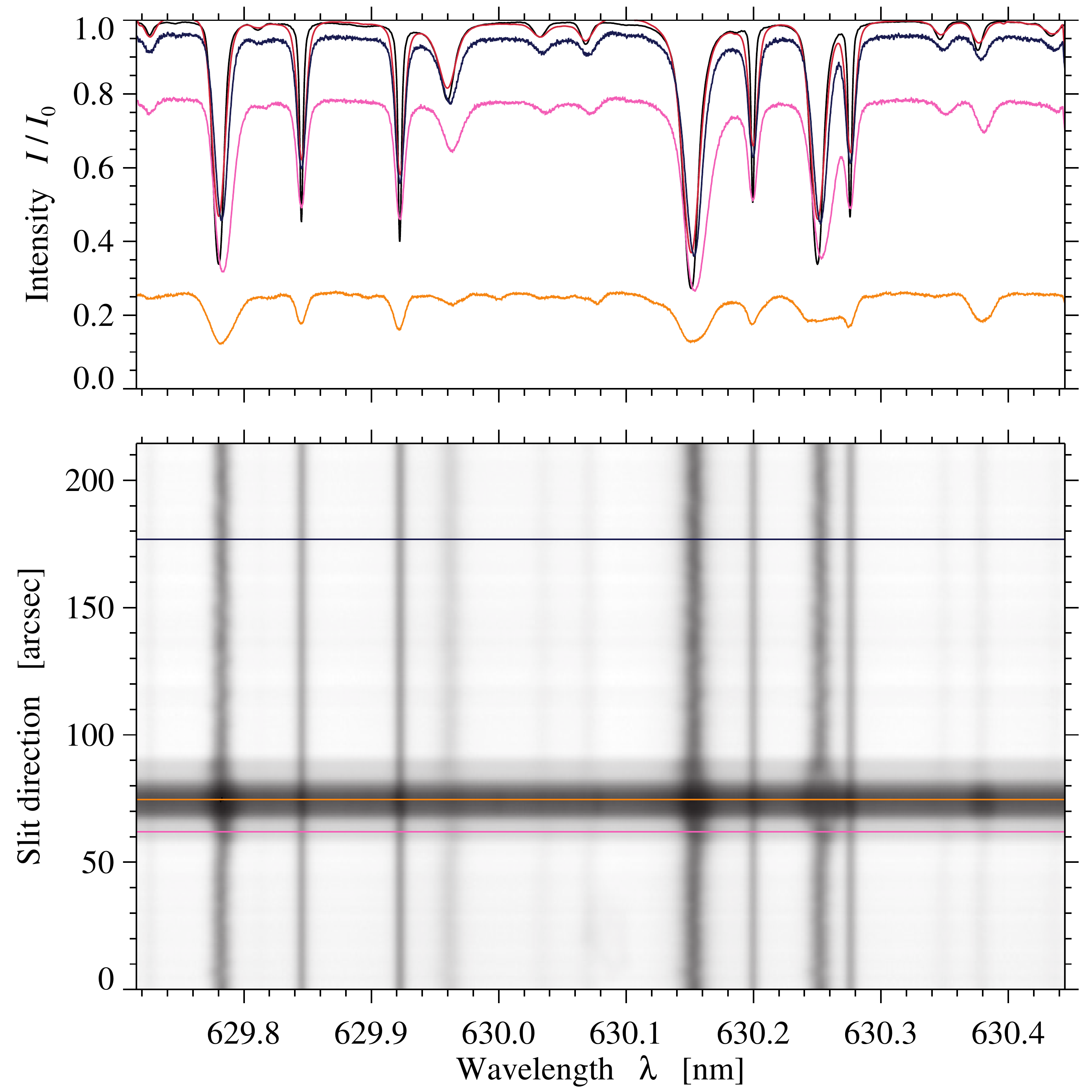}
\caption{Sunspot spectrum of the 630.2~nm spectral region (\textit{bottom}) 
    observed on 2016 April~29 and selected spectral profiles (\textit{top}): 
    FTS atlas (\textit{black}), average quiet-Sun (\textit{red}), quiet-Sun 
    (\textit{blue}), penumbra (\textit{orange}), and umbra (\textit{pink}).}
\label{FIG09}
\end{figure}

Figures~\ref{FIG08} and \ref{FIG09} illustrate the performance of telescope and 
spectrograph, depicting a Na\,D$_1$ 589.6~nm spectrum taken during the Mercury 
transit on 2016 May~9 and the spectral region around 630.2~nm, which contains 
two commonly used Fe\,\textsc{i} and nearby telluric lines. The sunspot spectrum 
acquired on 2016 April~29 (Fig.~\ref{FIG09}) clearly shows the Zeeman 
broadening of the Fe\,\textsc{i} lines, and a selection of quiet Sun, penumbral, 
and umbral spectral profiles is displayed in the upper panel. To demonstrate the 
very high resolving power of the spectrograph, a spectrum from the Kitt Peak 
Fourier Transform Spectral (FTS) atlas \citep{Neckel1984} is shown for 
comparison. The line widths are almost identical. However, the line depths of 
the photospheric lines are much deeper for the FTS spectrum because stray light 
contributions are negligible for this type of instrument, in contrast to the 
Einstein Tower spectra, where significant spectral stray light is present. The 
strong telluric lines in the observed spectrum are typical for low-altitude 
sites such as Potsdam.

\subsection{Photographic full-disk observations}

\begin{figure}[t]
\includegraphics[width=\columnwidth]{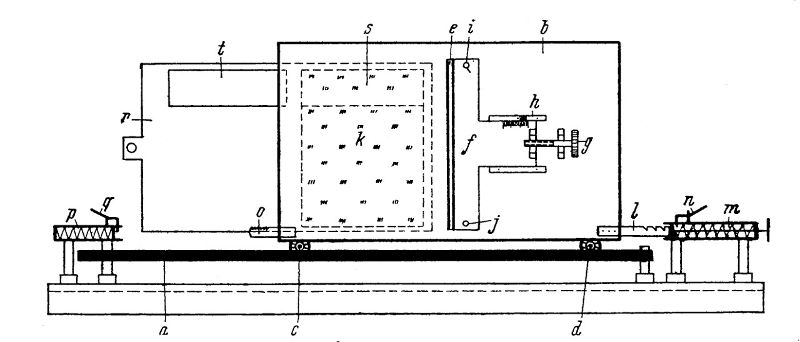}
\caption{Schematic sketch of the slit camera for solar full-disk images: ($a$)
    rail of polished round-bar steel, ($b$) movable rectangular cover plate, 
    ($c$,\,$d$) ball bearing mounted rolls, ($e$) slit, ($f$) adjustable 
    slit-jaw, ($g$) knurled-head screw for precise adjustment of the slit 
    width, ($h$) scale for the slit width, ($i$,\,$j$) tommy screws for 
    clamping the slit-jaw, ($k$) photographic plate behind the cover plate, 
    ($l$\,--\,$q$) spring-loaded mechanism to release the cover plate for an 
    exposure, and ($r$\,--\,$t$) devices to add intensity scale and directional 
    marks \citep[see Fig.~2 in][]{vonKlueber1948}.}
\label{FIG10}
\end{figure}

The solar image in the primary focus has a diameter of 
$d_{\scriptscriptstyle\bigodot} \approx 13$~cm so that direct imaging of the 
entire solar disk requires a detector of similar size, which is only feasible 
with photographic plates. For example, the largest monolithic CCD detectors 
embedded in PEPSI \citep{Strassmeier2015} have 112 megapixels and a detector 
size of 95~mm $\times$ 95~mm, which would be still too small for the full-disk 
image. Before CCD detectors became available, a photographic camera was required 
to uniformly illuminate a large-format photographic plate employing short 
($t_\mathrm{exp} \approx 10$~ms) exposure times to `freeze' the seeing. Iris 
blade shutters are impracticable because of the size of the solar image and the 
non-uniform illumination pattern. The design and principles of operation of a 
camera that overcomes these obstacles were described in \citet{vonKlueber1948}, 
and a schematic sketch of the photographic imaging device is shown in 
Fig.~\ref{FIG10}. This camera was in operation until the late 1980ies but does 
not any longer exist.

A highly polished steel rod with a diameter of 18~mm serves as the guide rail 
for the shutter. The rail has to be horizontal and dust free to deliver a 
uniformly exposed photographic plate. The shutter plate has a vertical slit, 
which can be adjusted in width to change the exposure time. The large-format 
photographic plates with dimensions of 18~cm $\times$ 24~cm are mounted in a 
wooden frame. A spring-loaded mechanism releases the shutter, which then slides 
rapidly ($v = $~1\,--\,2~m s$^{-1}$) past the photographic plate. To keep the 
speed constant, the approximately 1-meter long rail was tilted by about 2~mm to 
compensate friction. The length and tension of spring as well as the length of 
the holding and release mechanism are critical because the acceleration of the 
shutter has to stop before the slit reaches the photographic plate so that the 
motion is uniform. Finally, once the exposure is finished, a catch mechanism 
with a spring decelerates the shutter and arrests its motion. In a second 
exposure directional marks and a grayscale are added to determine the 
terrestrial coordinates and to facilitate photometric analysis of the solar 
full-disk images. Heliographic coordinates are computed with the help of the 
coelostat angle and the solar position angle.

\begin{figure*}[t]
\centering
\includegraphics[height=0.85\columnwidth]{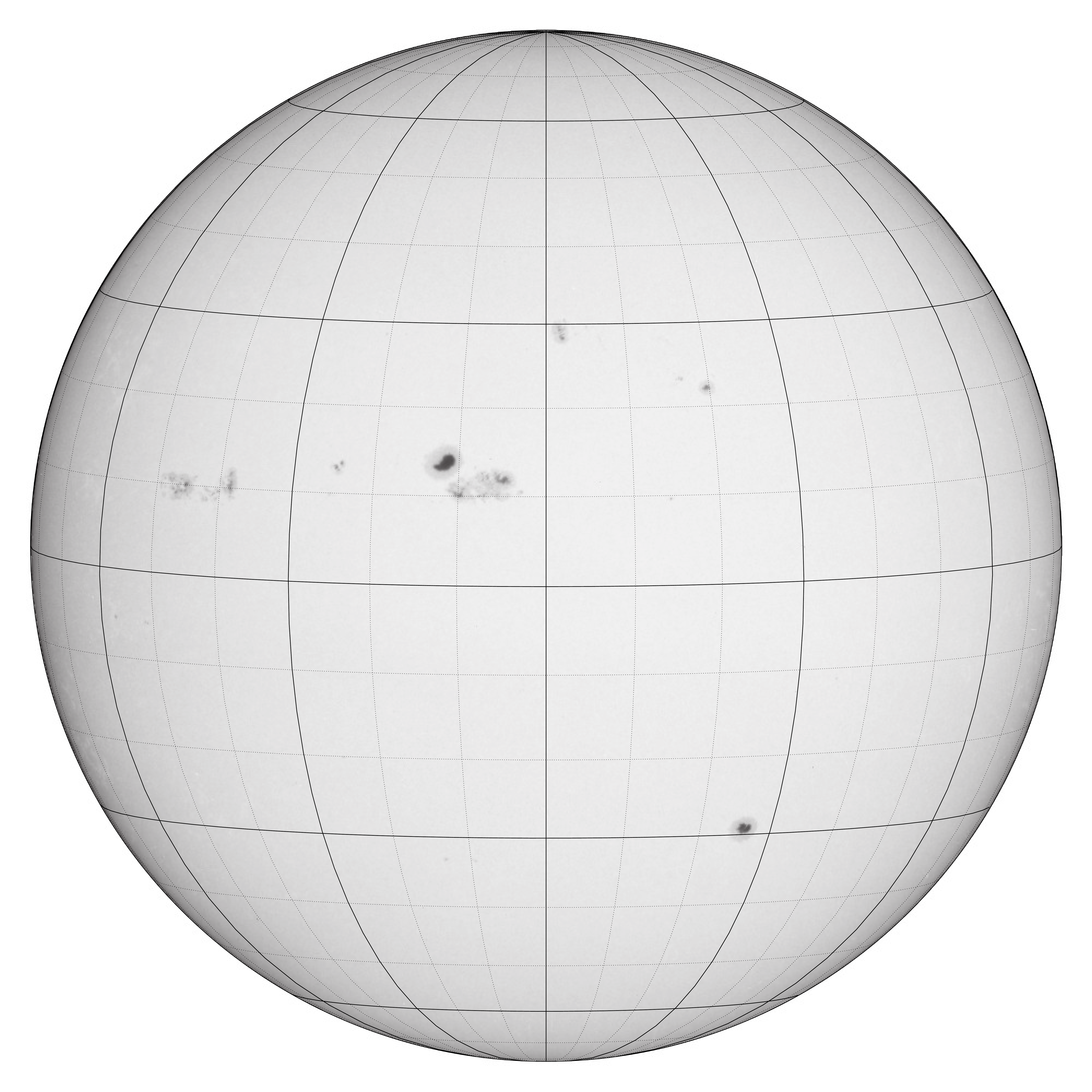}
\hspace*{5mm}
\includegraphics[height=0.85\columnwidth]{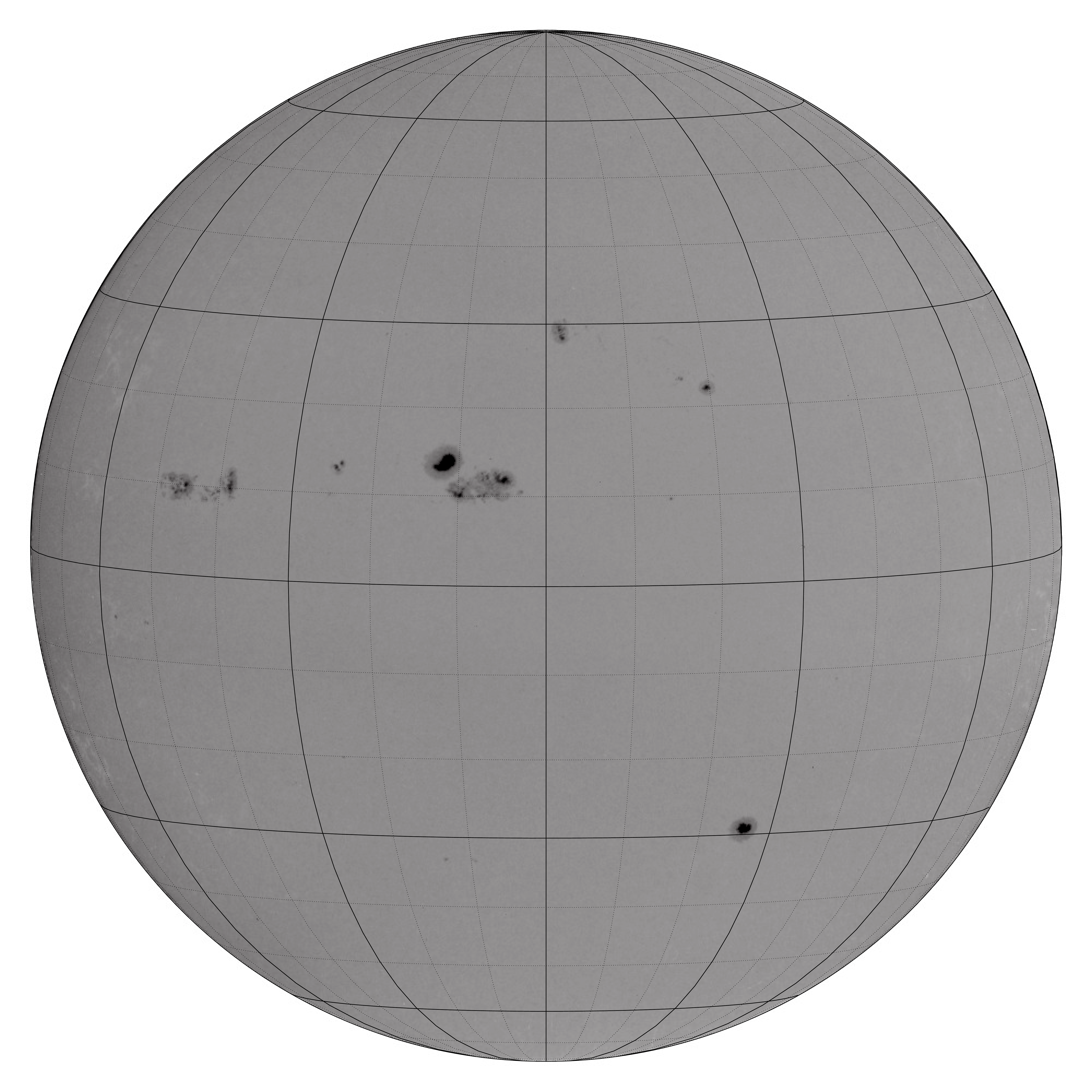}
\caption{Solar full-disk image (\textit{left}) and corresponding limb darkening 
    corrected image (\textit{right}) observed at the solar observatory 
    Einstein Tower at 07:16~UT on 1947 July~17. The position angle 
    $P=4.3^\circ$ is corrected, and heliographic coordinates are indicated in 
    intervals of $30^\circ$ (\textit{solid lines}) and $10^\circ$ 
    (\textit{dotted lines}), respectively.}
\label{FIG11}
\end{figure*}

About 3000 full-disk images exist in the photographic plate archive of the 
Einstein Tower covering the period from 1943 to the mid-1980ies. The coverage is 
usually good but drops significantly towards the end of this period. These 
plates were digitized in recent years with the help of high-school students 
carrying out internships at AIP. The scanned images have $7086 \times 7086$ 
pixels, and the raw data are saved as monochromatic 8-bit images in the Tagged 
Image File Format (TIFF). Calibrated and limb-darkening corrected full-disk 
images are saved as 4-megapixel files in the Flexible Image Transport System 
(FITS) format. Examples are depicted in Fig.~\ref{FIG11}, where the 
limb-darkening was removed using a similar procedure as in \citet{Denker1999a}. 
These contrast-enhanced images show besides sunspots and pores also clear 
signatures of facular regions near the solar limb with good seeing conditions 
\citep{vonKlueber1948}.

Solar full-disk images are no longer taken at the Einstein Tower. However, the 
large format of the Apogee Alta camera covers a field-of-view (FOV) of 
530\arcsec\ $\times$ 530\arcsec\ in the primary focus, which is sufficient to 
cover even major active regions. Large-format neutral density and color filters 
are used to significantly reduce the light level, while still guaranteeing short 
exposure times ($t_\mathrm{exp} \approx  100$~ms). The image scale of direct 
solar images is 0.18\arcsec\ pixel$^{-1}$, which means undersampling by a factor 
of two, if the full telescope aperture is used. However, considering the seeing 
conditions at Potsdam, achieving diffraction-limited observations even with 
image restoration techniques provides a formidable challenge. On the other hand, 
the observing conditions are sufficiently good to detect facular regions near 
the solar limb and sunspot fine-structure in solar full-disk images (see 
right panel of Fig.~\ref{FIG11}).

%
%

\section{Research and development}

Many research and development projects have benefited from the light-gathering 
power of the 60-cm diameter solar telescope and the stable thermal environment 
of the spectrograph chamber. A major advantage offered by the Einstein Tower is 
its availability for long-term measurements and equipment tests. Optimization 
of and adjustments to the instrument design can thus easily be made at the 
institute's nearby workshops. As a consequence, mature instruments can be 
delivered to world-class telescopes, where telescope time is at a premium.

Components of the Stokes-IQUV spectropolarimeter \citep{Strassmeier2003, 
Ilyin2011} for the Potsdam \'Echelle Polarimetric and Spectroscopic Instrument 
\citep[PEPSI,][]{Strassmeier2008a, Strassmeier2015} were tested in the optical 
laboratory of the Einstein Tower before delivering the spectropolarimeter to 
the Large Binocular Telescope (LBT) at Mt.\ Graham International Observatory in 
Arizona. PEPSI resides at the LBT's Gregorian focus station and achieves a 
spectral resolution of ${\cal R} \approx 120\,000$ in the polarimetric mode, 
which is available for the wavelength range 383\,--\,907~nm using three 
exposures for full spectral coverage. The light-gathering power of the $2 \times 
8.4$-meter aperture LBT enables a polarimetric accuracy of up to $10^{-4}$ for 
stellar observations previously only available for solar spectropolarimetry. The 
tested opto-mechanical components included rotary stages holding the 
polarimetric calibration optics, which consists of a Glan-Thompson prism and two 
quarter-wave plates. In addition, a modified Foster prism beam-splitter, serving 
as a linear polarizer, was evaluated.

\begin{figure}[t]
\centering
\includegraphics[height=0.49\columnwidth]{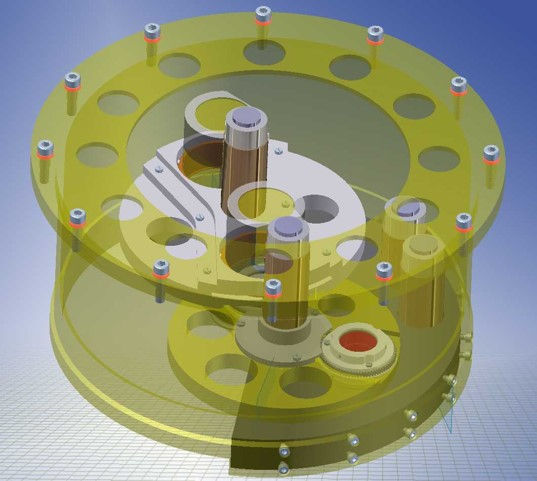}
\hfill
\includegraphics[height=0.49\columnwidth]{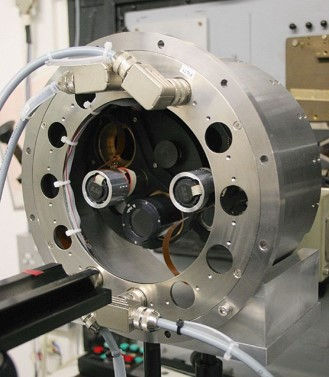}
\caption{Construction model of the GPU (\textit{left}) and the GPU mounted on
    the optical in front of the spectrograph entrance slit during calibration
    measurements (\textit{right}).}
\label{FIG12}
\end{figure}

The altitude-azimuth mount \citep{Volkmer2012} of the 1.5-meter aperture GREGOR 
solar telescope \citep{Schmidt2012} leads to a complex optical path traversing 
elevation and declination axes as well as the image rotator and the GREGOR 
Adaptive Optics System \citep[GAOS,][]{Berkefeld2012}. Cover windows of the 
partially evacuated light path and a total of 16 mirrors introduce a time 
dependent instrumental polarization. The GREGOR Polarimetric Unit 
\citep[GPU,][]{Balthasar2011, Hofmann2012} is located at the telescope's 
secondary focus, where the FOV is already significantly reduced, and the beam 
diameter is sufficiently small to accommodate the available size of the 
polarization calibration optics. The optical elements are mounted on two 
motorized wheels and motorized rotation stages are used to position a linear 
polarizer and two quarter wave retarders for the visible and near-infrared. The 
GPU is located still in the symmetric part of the optical train before the 
Nasmyth mirror, which directs the light into the coud\'e optical train. Thus, 
the instrumental polarization of the coud\'e optical train, image rotator, and 
AO system is calibrated. A construction drawing of the GPU is shown in the left 
panel of Fig.~\ref{FIG12}.

The characteristics of the polarizers and retarders were determined in the 
optical laboratory of the Einstein Tower \citep{Hofmann2009}, where the GPU was 
placed in front of the spectrograph entrance (see right panel of 
Fig.~\ref{FIG12}). An air-spaced Marple-Hess prism (a particular form of a 
double Glan prism) has to withstand an extreme power density of 25~W~cm$^{-2}$ 
and has to achieve an extinction ratio of $10^{-5}$\,--\,$10^{-6}$. The 
measurements showed that a retardation accuracy of $\lambda/100$ and a 
positioning accuracy of $0.1^\circ$ yields M\"uller matrices with an accuracy at 
the $10^{-4}$ level \citep{Hofmann2008}. The measured extinction ratio was $2 
\times 10^{-6}$, and the angular position of the intensity minimum was 
independent of the wavelength range (400\,--\,1000~nm) within the measurement 
errors. The fast axis of combined zero-order retarders is a function of 
wavelength, which was determined across the visible (379\,--\,980~nm) and 
near-infrared (656\,--\,1105~nm) wavelength ranges. Finally, the tests 
demonstrated that the broad- and narrow-band efficiency of the GPU to generate 
circular polarized light is very close to unity across the desired wavelength 
range and matches the design specifications \citep{Hofmann2009}.

\begin{figure}[t]
\centering
\includegraphics[height=\columnwidth]{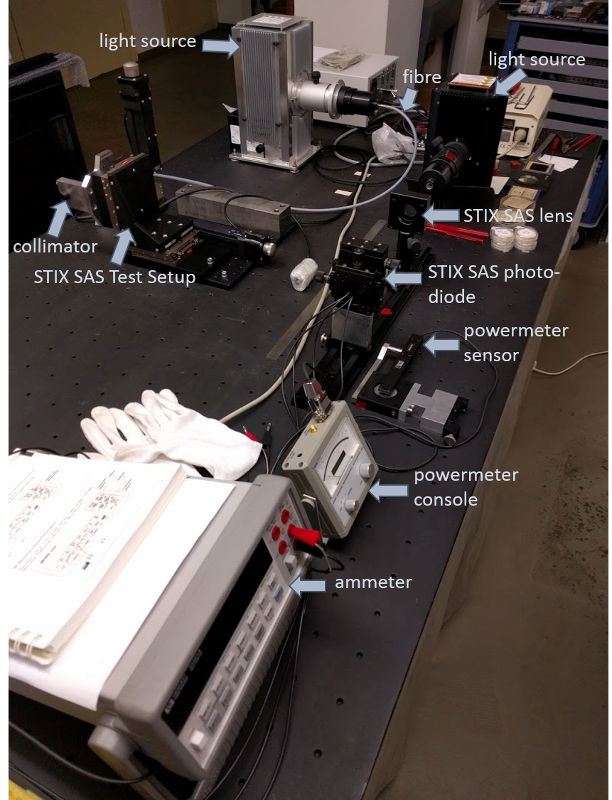}
\caption{Optical table with two set-ups for STIX: SAS simulator to calibrate 
    the STIX lens and photodiode (\textit{right}) and SAS verification system
    of STIX flight and flight spare models (\textit{left}).}
\label{FIG13}
\end{figure}

Solar Orbiter \citep{Mueller2013} is an ESA mission to be launched in October 
2018 and serves a platform for multiple instruments including the 
Spectrometer/Telescope for Imaging X-rays \citep[STIX,][]{Benz2012}. Initial 
tests, optimization, and calibration of the STIX Aspect System (SAS) were 
performed at the Einstein Tower before test and verification campaigns are 
carried out with the flight and flight spare models. To improve the pointing of 
the spacecraft, SAS functions as a solar limb detector and provides very 
accurate positioning for STIX (few seconds of arc). This facilitates coordinated 
observing campaigns with other ground-based or spaceborne instruments 
\citep{Krucker2013}. An SAS simulator will be installed in the optical 
laboratory to investigate and verify the SAS data obtained during all phases of 
the Solar Orbiter mission. Some of the equipment (see Fig.~\ref{FIG13}) was 
tested in the optical laboratory of the Einstein Tower.

Besides research and development in solar instrumentation, other projects have 
exploited the unique characteristics of the Einstein Tower's spectrograph optics 
and the stable thermal environment of the subterranean spectrograph chamber. 
Advanced cementing technology was tested for a 20-cm diameter open reflector 
with plane mirrors \citep{Neubert2012}, which will be used in laser 
communications with low earth orbit (LEO) satellites. Using the large, 35-cm 
diameter collimator lens of the spectrograph, the assembly and manufacturing 
accuracy was validated (nearing the 0.2\arcsec\ target), gravity-induced 
distortions in different mounting positions were identified, and long-term 
drifts were measured for the triple mirror assembly.

%
%

\section{Education and public outreach}

The Science Park Albert Einstein with its ensemble of historical science 
buildings in an English landscape garden is open to the public during daytime 
hours. It is home to the German Research Center for Geosciences (GFZ) and the 
Potsdam Institute for Climate Impact Research (PIK). In addition, remote 
stations of the Alfred Wegener Institute for Polar and Marine Research (AWI), 
the German Meteorological Office (DWD), and AIP are located at the 
Telegrafenberg. The annual ``Long Night of the Science'' is the main public 
outreach event, which attracts several thousands of visitors from the 
Berlin-Potsdam metropolitan area. Many smaller events throughout the year and 
guided tours for school classes and university courses complement this main 
event. Solar observations with the Einstein Tower are part of the astronomy and 
astrophysics curriculum for master students at the University of Potsdam. 
Experiments include measuring the magnetic field of sunspots and determining the 
solar differential rotation. Training in spectropolarimetric techniques is also 
offered to junior scientists at AIP, who typically do not have direct access to 
and hands-on experience with instruments at today's major astronomical research 
infrastructures.

%
%

\section{Conclusions}

First observations started at the Einstein Tower in 1924 December~6. More than 
90 years later the telescope and its instrumentation are still an important 
asset in AIP's research portfolio. Nowadays, activities are focused on (1) 
education, training, and public outreach and (2) research and development. 
Gratifyingly, the upgraded CCD camera system and investments improving 
positioning of the telescope and tracking active regions will facilitate new 
science observations. With the second generation instrumentation of GREGOR and 
the European Solar Telescope \citep[EST,][]{Collados2010a} on the horizon, we 
expect that the Einstein Tower will play an important role in the related 
research and development activities.

%
%

\acknowledgements SJGM is grateful for financial support from the Leibniz 
Graduate School for Quantitative Spectroscopy in Astrophysics, a joint project 
of the Leibniz Institute for Astrophysics Potsdam and the Institute of Physics 
and Astronomy of the University of Potsdam. The major restoration of the 
Einstein Tower in the years 1997\,--\,1999 was made possible by generous funds 
of the W\"ustenrot Foundation. Ongoing conservation and maintenance of the 
Einstein Tower is supported by the State of Brandenburg through institutional 
and extraordinary funding. The Einstein Tower received in 2015 a planing and 
conservation grant of the Getty Foundation's ``Keeping It Modern'' initiative 
exploring options for preserving the complex and challenging building structure.



\end{document}